# Complete Human Mitochondrial Genome Construction Using L-systems

Sk. Sarif Hassan[a,c], Pabitra Pal Choudhury[a], Amita Pal[b], R. L. Brahmachary[c] and Arunava Goswami[c]

[a]Applied Statistics Unit, Indian Statistical Institute, 203 B. T. Road, Calcutta, 700108 India. pabitrapalchoudhury@gmail.com [P. P. C.]; sarimif@gmail.com [S. S. H.];

[b] Bayesian Interdisciplinary Research Unit (BIRU), Indian Statistical Institute, 203 B. T. Road, Calcutta, 700108 India. pamita@isical.ac.in [A. P.] and

[c]Biological Sciences Division, Indian Statistical Institute, 203 B. T. Road, Calcutta, 700108 India. agoswami@isical.ac.in [A.G.].

**Keywords**
- Mitochondrial genome
- L-system
- Genomics
- Assembly Method

**Footnotes**
- To whom correspondence should be addressed. E-mail: agoswami@isical.ac.in / sarimif@gmail.com
- *Author contributions*: A. G. conceptualized the experiments and performed entire research with S. S. H.; A. G. and S. S. H. acknowledges A. P., P. P. C. and R. L. B. and Ms. Rebecca French (Grad student of Prof. Michael Hochella) of Virginia Tech, USA for their enormous help while doing these series of experiments.
- Conflict of interest statement: The authors declare no conflict of interest.

**Abstract:** Recently, scientists from The Craig J. Venter Institute reported construction of very long DNA molecules using a variety of experimental procedures adopting a number of working hypotheses. Finding a mathematical rule for generation of such a long sequence would revolutionize our thinking on various advanced areas of biology, viz. evolution of long DNA chains in chromosomes, reasons for existence of long stretches of non-coding regions as well as would usher automated methods for long DNA chains preparation for chromosome engineering. However, this mathematical principle must have room for editing / correcting DNA sequences locally in those areas of genomes where mutation and / or DNA polymerase has introduced errors over millions of years. In this paper, we report the basics and applications of L-system (a mathematical principle) which could answer all the aforesaid issues. At the end, we present the whole human mitochondrial genome which has been generated using this mathematical principle using PC computation power. We can claim now that we can make any stretch of DNA, be it 936 bp of olfactory receptor, with or without introns, mitochondrial DNA to $3 \times 10^9$ bp DNA sequences of the whole human genome with even a PC computation power.

## 1. Introduction:

Earlier, in [1, 2] we described and designed the human olfactory receptor gene OR1D2 using one L system [3]. Actually in an experiment we had taken OR1D2, OR1D4 and OR1D5, three full length olfactory receptors present in an olfactory locus in human genome. These receptors are more than 80% identical in DNA sequences and have 108 base pair mismatches among them. However, we were successful to find a mathematical rule in those mismatches. We find an L-system generated sequence which can be inserted into the OR1D2 subfamily specific star model and novel full length olfactory receptors can be generated.

Gibson DG et al. (2008) [4] from The Craig J. Venter Institute in USA surprised us in a paper by preparing whole genome of *Mycoplasma genitalium* within yeast cells where he and his colleagues could stitch 25 overlapping DNA fragments to form a complete synthetic genome (524 kb). This was a phenomenal discovery. Gibson et al (2009) [5] in a paper published in Nature Methods showed long DNA chain could made easily using an elegant experimental method in which concerted action of a 5' exonuclease, a DNA polymerase and a DNA ligase lead to a thermodynamically favored isothermal single reaction. In the beginning, researchers recessed DNA fragments and this process yielded single-stranded DNA overhangs that specifically annealed, and then covalently joined them. In this process, they could assemble multiple overlapping DNA molecules and surely enough mechanism of action behind making a full chromosome is now ready. But initially to organize *assembly method* proposed by Gibson DG et al. (2008) four DNA cassettes of 6-kb (which could be had from Yeast genome itself) are needed. But what governs the whole yeast genome? In this paper, we are trying to explore a possible underlying mathematical principle for making whole genome.

After the preliminary success in [1], we are sufficiently enthused to take up the whole human mitochondrial DNA which is equivalent to any bacterial DNA for the purpose of designing/constructing with the help of a set of L-systems. We have a strong belief that nature might use one nucleotide to start with in order to construct the whole chromosome and finally the genome. On the basis of this conjecture, we are motivated to pick up the L- system methodology.

We claim that the proposed methodology could be used in an automated system to seamlessly construct synthetic and natural genes, for modulation of genetic pathways and finally entire genomes, and could be a useful chromosome engineering tool.

## 2. Design of Mitochondrial DNA

### 2.1 Construction Methodology of L Systems:

How we are going to select the set of L systems [2] is a relevant question; let us brief the corresponding algorithm as follows:

The design of L systems are as follows where axiom (starting symbol) for the L system is A.
The nucleotide A produces first four consecutive base pairs of the mitochondrial DNA and C, T and G produce next consecutive four base pairs respectively.

Now if the number of mismatches of the DNA sequence is less than three then an L system could be chosen as: the nucleotide A produces the remaining mismatches and C, T and G all produce A.

But if the numbers of mismatches occur in between two and fifteen then an L-system could be chosen as follows: the nucleotide C produces first one third of the remaining mismatches, T produces next one third,

G produces the remaining and finally A produces the CTG to achieve the remaining mismatches in the 2[nd] iteration of the L-system.

Based on the proposed methodology as stated above, we go on generating the L-system iterations until it crosses the given mitochondrial length. Now we compare the generated sequence with the given mitochondrial sequence and mismatching portions are again tried by another set of L-systems following the above pick up policy. And the same process is continued until the whole mitochondrial sequence is matched.

In this way we have the following twenty four L-systems' covering the mitochondrial sequence as shown in the as given below:

**Set of L systems:**

```
L-System for iteration number 1:
A --> GATC
C --> ACAG
T --> GTCT
G --> ATCA
______________________________

L-System for iteration number 2:
A --> TAGG
C --> TCAG
T --> TGGA
G --> TATT
______________________________

L-System for iteration number 3:
A --> AGCA
C --> GGGA
T --> TTGT
G --> GGCC
______________________________

L-System for iteration number 4:
A --> GATT
C --> TGCC
T --> CGTC
G --> AACT
______________________________

L-System for iteration number 5:
A --> TGCC
C --> CTAA
T --> CTAC
G --> CCAG
______________________________

L-System for iteration number 6:
A --> TGCC
C --> ACTA
T --> CCAG
G --> ACGG
```

```
________________________
L-System for iteration number 7:
A --> TGCC
C --> CCAG
T --> ACGT
G --> CTAC
________________________

L-System for iteration number 8:
A --> TGCC
C --> CGAG
T --> TTCG
G --> ATTA
________________________

L-System for iteration number 9:
A --> GCCG
C --> ATTC
T --> AGAG
G --> AATA
________________________

L-System for iteration number 10:
A --> GCCG
C --> GAGG
T --> GCTG
G --> AGGT
________________________

L-System for iteration number 11:
A --> GCCG
C --> CTTA
T --> CTAG
G --> TGTG
________________________

L-System for iteration number 12:
A --> GCCG
C --> TATG
T --> TGAT
G --> ATTA
________________________

L-System for iteration number 13:
A --> ATGA
C --> TTAA
T --> CATA
G --> AATC
________________________

L-System for iteration number 14:
A --> TTAC
C --> AATC
```

```
T --> ACCT
G --> CCAA
________________________

L-System for iteration number 15:
A --> CCTC
C --> AACC
T --> TCCT
G --> CTTA
________________________

L-System for iteration number 16:
A --> CCTC
C --> TTAG
T --> TTGC
G --> GCGC
________________________

L-System for iteration number 17:
A --> CCTA
C --> GTTG
T --> CGCC
G --> GTGA
________________________

L-System for iteration number 18:
A --> CCTG
C --> TTGC
T --> GCCG
G --> TGAA
________________________

L-System for iteration number 19:
A --> CCTG
C --> GCCG
T --> GATA
G --> ATAC
________________________

L-System for iteration number 20:
A --> CCTG
C --> ATAT
T --> GAAA
G --> GTAT
________________________

L-System for iteration number 21:
A --> ATTG
C --> AAAT
T --> TGAG
G --> GTAT
________________________

L-System for iteration number 22:
```

```
A --> CTG
C --> AAT
T --> GAA
G --> TGA
```
___________________________

**2.2 Remark**

We have observed in the publicly available database (from NCBI) the human mitochondrial DNA sequence consist one 'n'. And just before this 'n' there is a nucleotide C which allows to replace the 'n' in (3061 position; http://www.ncbi.nlm.nih.gov/nuccore/NC_012920) by any one of four nucleotides. The interesting fact is that when 'n' is replaced by A or T to cover up the whole mitochondrial sequence, it takes only 22 L-systems whereas 24 L-systems if 'n' is replaced by C or G.

**3. Conclusion and future efforts:**

In the near future with the same methodology we could proceed to design/construct the chromosome level and hence in due course the whole human chromosome might be designed. Results from the above mentioned three studies clearly show that coding regions of the genes in the genome is made from the specificities of the particular L-system and the large amount of non-coding region is necessary for giving the stability of the strings of DNA on thermodynamic scale. Another important point emanated from these studies is that application of L-systems in a particular combination allows repairing of sudden change in genome locally and thereby helps to keep the string of chromosomal DNA intact. This result can be further extended to study of the nuclear chromosomal DNA. Also results from this study show clearly that this kind of approach can be very useful for correcting the ambiguity of the DNA sequences present in published and genomic sequences of various genbank databases.

**Acknowledgements**

This work was supported by Department of Biotechnology (DBT), Govt. of India grants (BT/PR9050/NNT/28/21/2007 & BT/PR8931/NNT/28/07/2007 to A. Goswami) & NAIP-ICAR-World Bank grant (Comp-4/C3004/2008-09; Project leader: A. Goswami) and ISI plan projects for 2001-2011. Authors are grateful to their visiting students Rajneesh Singh and Snigdha Das for their technical help in making advanced C programs and other computer applications on Windows support used for this study.